\documentclass[useAMS,usenatbib]{mn2e}
\usepackage{subfigure}
\usepackage{setspace}
\usepackage{graphicx}
\def\spose#1{\hbox to 0pt{#1\hss}}
\def\simlt{\mathrel{\spose{\lower 3pt\hbox{$\mathchar"218$}}
     \raise 2.0pt\hbox{$\mathchar"13C$}}}
\def\simgt{\mathrel{\spose{\lower 3pt\hbox{$\mathchar"218$}}
     \raise 2.0pt\hbox{$\mathchar"13E$}}}

\title[A Simulated Polar Disc Galaxy]{The Halo Shape and Evolution  of Polar Disc 
Galaxies}
\author[Snaith et~al.]{\parbox{\textwidth}
{O.N. Snaith$^{1,2,3}$, 
B.K. Gibson$^{1,4,5}$, 
C.B. Brook$^{1,6}$, 
A. Knebe$^{6}$, 
R.J. Thacker$^{4}$, 
T. R. Quinn$^{7}$,
F. Governato$^{7}$
and P. B. Tissera$^{2}$}\vspace{0.4cm}\\
\parbox{\textwidth}{$^1$Jeremiah Horrocks Institute,
University of Central Lancashire, Preston, PR1~2HE, UK\\
$^2$Instituto de Astronom\'ia y F\'isica del Espacio, Conicet-UBA, CC67, 
Suc. 28, Ciudad de Buenos Aires, Argentina\\
$^3$GEPI, Observatoire de Paris, CNRS, Univ. Paris Diderot, 5 Place Jules Janssen, F-92195 Meudon, France \\
$^4$Department of Astronomy \& Physics, Saint Mary's University, Halifax,
Nova Scotia, B3H~3C3, Canada\\
$^5$Monash Centre for Astrophysics, School of Mathematical Sciences,
Monash University, Clayton, VIC, 3800, Australia\\
$^6$Grupo de Astrofi\'isica, Depatamento de Fisica Teorica, Universidad
Aut\'onoma de Madrid, Cantoblanco, E-280049, Spain\\
$^7$Department of Astronomy, University of Washington, Seattle, WA 98195, USA}}

\begin{document}
\date{\today}
\pagerange{\pageref{firstpage}--\pageref{lastpage}} \pubyear{2011}
\maketitle
\label{firstpage}

\begin{abstract}
We examine the properties and evolution of a simulated polar disc galaxy. This
galaxy is comprised of two orthogonal discs, one of which contains old stars
(old stellar disc), and the other, containing both younger stars and the cold
gas (polar disc) of the galaxy. By exploring the shape of the inner region of
the dark matter halo, we are able to confirm that the halo shape is a oblate
ellipsoid flattened in the direction of the polar disc. We also note that there
is a twist in the shape profile, where the innermost 3~kpc of the halo flattens
in the direction perpendicular to the old disc, and then aligns with the polar
disc out until the virial radius. This result is then compared to the halo shape
inferred from the circular velocities of the two discs. We also use the temporal
information of the simulation to track the system's evolution, and identify the
processes which give rise to this unusual galaxy type. We confirm the proposal
that the polar disc galaxy is the result of the last major merger, where the
angular moment of the interaction is orthogonal to the angle of the infalling
gas. This merger is followed by the resumption of coherent gas infall. We
emphasis that the disc is rapidly restored after the major merger and that after
this event the galaxy begins to tilt. A significant proportion of the infalling
gas comes from filaments. This infalling gas from the filament gives the gas its
angular momentum, and, in the case of the polar disc galaxy, the direction of the
gas filament does not change before or after the last major merger.  

\end{abstract}


\section{Introduction}

Classical disc galaxies consist of flattened distributions of stars, gas, and
dust, co-rotating with the angular momentum vectors more or less
aligned;\footnote{In the presence of warps though, there may be some
misalignment driven by external torques \citep{Roskar2010}.} polar disc
galaxies\footnote{Originally called `multi-spin galaxies' by \citet{Rubin1994}.}
are significantly more rare, possessing two orthogonal discs. An original survey
of galaxies consisting of orthogonal structures of stars was carried out by
\citet{Whitmore1990} who identified a sample of polar {\it ring} galaxies. The
`primary' discs in such systems are typically classified as lenticulars (S0), or
Ellipticals, with a second, younger, structure aligned with angular momentum
axis of the primary disc. Fewer than 1\% of S0 galaxies contain a polar
structure. 

Subsequent studies of NGC4650A revealed a subclass of structures where the
orthogonal distribution of stars was not a {\it ring} but a second  fully-formed
continuous {\it disc} structure, \citep{Iodice2002, Gallagher2002,Swaters2003,
Maccio2006,  Brook2008}. Systematic studies suggest that their stellar
populations, light distribution, and gas-phase characteristics, are not
dissimilar from those of classical discs - e.g., exponential light profiles
\citep{Schweizer1983} and \textsc{HI} mass-to-optical luminosities typical of
mass to B-band luminosity typical of late-type spirals \citep{Huchtmeier1997,
Arnaboldi1997, Sparke2000, Cox2006}. 

Thus there exist {\it polar ring} galaxies as well as {\it polar disc} galaxies. 
We will discuss the latter type in the following paper. On of the important uses of
polar disc galaxies is a direct result of having orthogonal discs at the bottom
of a halo potential well. One stellar disc historically allowed astronomers to
calculate the existence of dark matter halos, (see \citet{Begeman1991} for a
detailed discussion). \citet{Iodice2003} note that the existence of two
orthogonal discs means that the shape of the inner region of the halo can be
recovered from observations.  

There are several theories proposed to explain the origin of polar ring and polar disc 
galaxies:

\begin{enumerate}
\item {\bf Mergers} : As envisioned by \citet{Bekki1997, Bekki1998} and 
\citet{Bournaud2003}, polar rings can eventuate for very specific 
collision configurations where the collision is `head-on' and the 
initial angular momentum low.
\item {\bf Tidal Accretion} : Proposed by \citet{Schweizer1983} and 
simulated by \citet{Reshetnikov1997}, this scenario involves the capture 
of a gas-rich secondary by the (future) polar disc host halo.  The 
interaction-induced polar ring is formed more readily than in the 
aforementioned merger-induced scenario \citep{Bournaud2003, Combes2006}.  
\item {\bf Infall from Filaments} : If gas falls into a galaxy along 
cosmic filaments that are inclined to the stellar disc, a polar disc may 
form \citep{Maccio2006} \& \citep{Brook2008}. 
\citet{Combes2006} and \citet{Bournaud2003} suggest such cosmological 
infall is a viable alternative to tidal accretion for the NGC~4650A 
system e.g. \citep{Iodice2006, Spavone2010, Spavone2011}.
\item {\bf Resonance} : This approach assumes that the polar disc is 
formed via resonant coupling to a triaxial halo potential 
\citep{Tremaine2000}. In this picture, a disc is taken to lie in the 
plane of symmetry of the triaxial halo and the halo tumbles with respect 
to the disc. As the tumbling slows, the stars in the disc can get 
trapped in resonance with the halo. A stellar orbit inclined to the disc 
precesses at a slow retrograde rate and the star can be pushed into a 
polar orbit. This method has been invoked to best explain equal mass 
stellar disc systems.
\end{enumerate}

A seminal review of the field is provided by \citet{Combes2006}, to 
which all interested readers should refer for a rich survey of the 
field. The filamentary infall scenario (iii, above) of 
\citet{Maccio2006} has gathered momentum over the past few years as 
being the favoured origin of polar disc systems.  This can be traced, in 
part, to the fact that the alternatives do not account fully for the 
range of observations -- in particular, those which show that the polar 
disc is often of high mass, possesses extended and continuous structure, 
and lacks obvious evidence of starbursts that might be associated with 
polar structure formation. Scenario (i) can form  a polar disc/ring
galaxies when the relative initial velocities are low (e.g. \citet{Bournaud2003} ) 
and by (ii) in loose groups of galaxies such as UGC9796 \citep{Spavone2011}. 
Polar disc/ring galaxies are found in low density environments as the polar
structure is destroyed by mergers \citep{Maccio2006}. This suggests that 
each type of formation may exist in the universe.

Closely related to \citet{Maccio2006} scenario (iii, above), 
\citet{Brook2008} suggest that the polar disc forms as a result of the 
last major merger changing the angular momentum of the stars and gas of 
the stellar disc, and then subsequent gas continues to fall in along the 
old trajectory and builds up the polar disc.\footnote{Alternatively, the 
gas falling along filaments could change direction.} The basic 
observable characteristics of their simulated polar disc system were 
presented, including star formation rates, circular velocity profiles, 
and structural parameters of the discs, along with the aforementioned 
putative origin scenario.  The exact mechanism by which the dark halo 
aligns to the orthogonal discs and how the new infalling gas becomes 
misaligned to the orthogonal discs was {\it not} \rm 
studied by \citet{Brook2008}.

In what follows, we rectify this by examining the underlying physics 
governing the misalignment of the discs in the \citet{Brook2008} 
simulation, tracing its temporal evolution and association with 
large-scale structure, in order to put their origin scenario to the 
test, quantitatively.  The simulation and its basic properties are 
reviewed in \S~2 and \S~3, while the time evolution of the host halo's 
shape is detailed in \S~4.  The metallicity and age gradients of the 
discs are then confronted with recent observational work (\S~5), while 
the time evolution of the dark halo, gas, and stellar angular momenta is 
derived in \S~6.  Our conclusions and suggestions for future directions 
are made in \S~7.

\section{Code}

The polar disc system under examination here was realised with the 
gravitational N-body + smoothed particle hydrodynamics (SPH) code 
\textsc{Gasoline} \citep{Wadsley2004}. Gravitational forces are computed 
using a \citet{Barnes1986} tree-based algorithm; the code is spatially 
and temporally adaptive and employs periodic boundary conditions.  SPH 
is used to discretise the astrophysical gas and accounts for radiative 
cooling, ionisation, and excitation of the gas. The ultraviolet 
background radiation field is included, after \citet{Haardt1996}. The 
internal energy of the gas is calculated to conserve energy better than 
entropy and shocks are accounted for via artificial viscosity.

The star formation recipe employed within \textsc{Gasoline} is detailed 
by \citet{Stinson2006} and includes (i) a \citet{Kroupa2001} initial 
mass function (IMF), (ii) the effects of both Type~Ia and Type~II 
supernovae (neglecting the instantaneous recycling approximation), and 
(iii) a blast-wave feedback scheme with which to couple supernova energy 
to the surrounding interstellar medium.

For this simulation, first introduced by \citet{Brook2008}, volume 
re-normalisation was employed to ensure high spatial and force resolution 
in the area of interest, while self-consistently tracking the 
gravitational torques imparted by large-scale cosmological structure. 
The cosmological framework was a concordant $\Lambda$CDM, with 
$\Omega_0$=0.3, $\Omega_\Lambda$=0.7, $\sigma_8$=0.81, $h$=0.7, where 
$\Omega_0$ is the matter density, $\Omega_\Lambda$ is the fraction of 
the energy density of the Universe due to the cosmological constant, 
$\sigma_8$ is the rms density fluctuation in 8~Mpc sub-samples, and $h$ 
the Hubble Constant. The initial power spectrum was calculated using 
\textsc{Cmbfast} \citep{Zald1998,Seljak1996}. 

An initial, coarse, dark matter only simulation was run to redshift 
$z$=0 in a 28.5$h^{-1}$~Mpc box; this is large enough to provide 
realistic tidal torques with sufficient resolution for halo 
identification. A host halo was selected using \textsc{AHF} 
\citep{Gill2004, Knollmann2009}, and the particles therein traced back 
to their respective locations in the initial conditions. The volume is 
then re-centred on this region and said region `re-populated' with lower 
mass dark matter particles in concentric shells of increasing 
resolution. Each `generation' of dark matter particles is 1/8 the mass 
of the previous generation. The highest resolution dark matter particles 
have a mass $9.5 \times 10^{4}$~M$_\odot$. Only within the high 
resolution region are gas particles now placed, and tracked forward in 
time in a re-simulation of the volume, now including the effects of star 
formation and feedback. These gas particles each have a mass $1.6 \times 
10^{4}$~M$_\odot$; star particles which form from these gas particles 
have masses of $3.3 \times 10^{3}$~M$_\odot$. The force resolution 
within the inner, high-resolution, region is 150~pc.  At redshift $z$=0, 
there are $3.5 \times 10^{6}$ baryonic particles within the inner 
20~kpc.

\section{Basic Characteristics of the Simulated System}

\begin{figure}
\centering
\includegraphics[scale=0.4,clip]{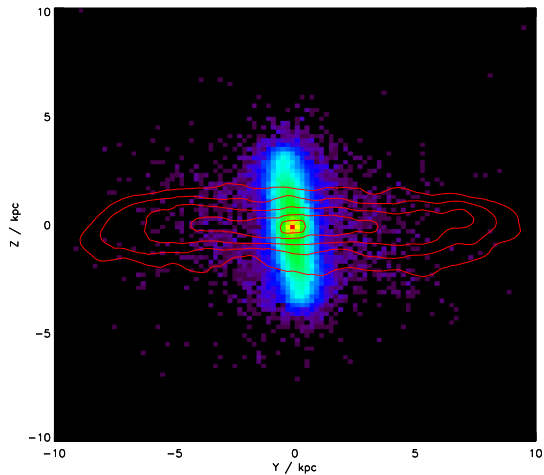}
\caption{Schematic density distributions of the intermediate-age stars
associated with the primary stellar disk (colours; vertical in this plane)
and cold gas associated with the secondary polar disk (contours; 
horizontal in this plane).}
\label{Fig:galaxyimage} 
\end{figure}

As noted earlier, the simulation discussed here was first introduced by 
Brook et~al. (2008), but a detailed examination of its dark matter halo 
characteristics, relationship to large scale structure, stellar and 
gas-phase metallicity distributions, and their relationship to the 
system's proposed formation, was reserved for future work.  We address 
these points in what follows.  A schematic of the system is shown in 
Fig.~\ref{Fig:galaxyimage} where the older, `primary', stellar disc is 
shown in colour (vertical in this figure) and the cold gas associated 
with the younger, 'secondary', polar disc is represented by the red 
contours (horizontal in this figure).

We can identify the polar and stellar discs cleanly via the ages of 
their associated star particles : recently-formed stars  ($t>$9~Gyrs) 
define the `polar disc', while intermediate-age stars 
(4$<t<$6~Gyrs) define the `stellar disc'.  This simple delineation 
is used to define the disc orientations for our subsequent analyses.  
Stars formed early ($t<$4~Gyrs) are associated with the halo of the 
galaxy and can be traced to the stars which formed before or during the 
last major merger.  

The system as a whole has a dynamical (stellar) mass of $1.6 \times 
10^{11}$~M$_\odot$ ($1.0 \times 10^{10}$~M$_\odot$) within its virial 
radius ($\sim$130~kpc) at z=0.17. The stellar mass of the old stellar disc
is $1.9 \times 10^{9}$~M$_\odot$ while the polar disc has a mass of $9.8 \times 
10^{8}$~M$_\odot$ -- i.e., the polar disc has half the stellar mass of the 
stellar disc.  The gas mass in the polar disc is approximately the same 
as the stellar mass at z=0.17, so the baryonic mass in the two 
discs is approximately the same. Both discs are less massive than the very old spheroid, 
(with stars formed at less than 4 Gyr which has a mass of  $6.2 \times 
10^{9}$~M$_\odot$). Of the three polar ring galaxies studied in Spavone et 
al. (2010, 2011), two show a polar disc more massive than the host galaxy, 
while UGC 7576 has a polar \emph{ring} 2.7 times less massive than the host galaxy. 
Our polar disc is 3.6 times less massive than the host, a value similar to that found for
UGC 9796 by \citet{Spavone2011}
The similarity in the masses of the two discs differs from NGC  4650A \citep{Spavone2011}
These mass values are highly dependant on the age 
cuts used to define the disc. The original values from \citet{Brook2008} give 
significantly different mass values for the two discs and so the relative 
masses of the two components is mostly dependant on how one chooses to define 
it, rather than an intrinsic property. For example, if we decompose using the 
method of \citet{Abadi2003}\footnote{This galaxy is decomposed according to 
the ratio of $\epsilon=J_z(E)/J_{circ}(E)$, where $J_z(E)$ is the z component of 
the angular momentum of the galaxy and $J_{circ}(E)$ is the angular momentum of 
the circular orbit with the same energy E. Disc stars are chosen via a cut in 
$\epsilon$ close to 1.} then the stellar masses of each disc are the same, so 
that the total baryonic mass of the polar disc is twice that of the stellar disc.

\begin{figure}
\centering
\includegraphics[scale=0.4,clip]{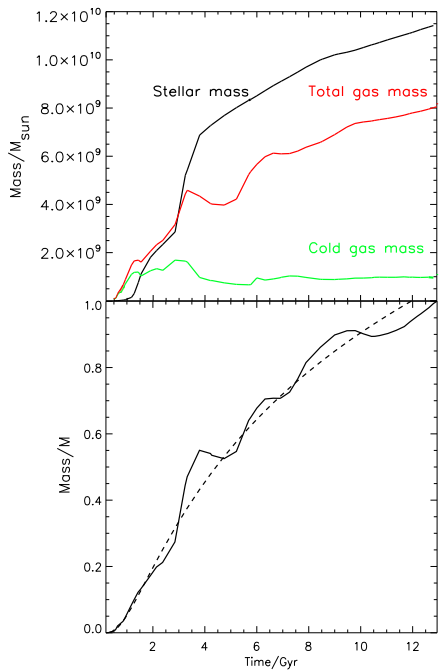}
\caption{The upper panel shows the mass accretion of the baryonic components, 
with galaxy mass against the age of the universe. The bottom panel shows the 
dark matter accretion history, where the dotted line is the best fit \citep{Wechsler2002} function.}

\label{fig:accrehist} 
\end{figure}

The stellar disc is essentially gas-free by redshift 
$z$$\sim$0.2, but prior to this, there is an extended period of time 
during which there is gas in both discs.  The scale-lengths of the 
stellar and polar discs are 1.0 and 3.4~kpc, respectively, with both 
showing disc truncations near $\sim$3 disc scale-lengths. The two discs 
are nearly orthogonal, with an inclination of $\sim$84$^\circ$ between 
the two.  The circular velocity at 2.2 scale-lengths for both discs is 
$\sim$110~km/s, based on the \citet{Brook2008} analysis. The polar 
disc system reported by \citet{Maccio2006} 
possesses similar masses and radial extents for its two discs.  We show 
the mass accretion history of the galaxy in Fig. \ref{fig:accrehist}.
We can, in some sense, think of this history in terms of three
epochs of `star formation' : a rapid initial phase ($t$$<$4~Gyrs), 
an intermediate phase (4$<$$t$$<$8~Gyrs), and a more gentle, later, 
phase ($t$$>$8~Gyrs). 

\subsection{History of Formation}

The primary disc first forms at redshift $z$=3.3 (age=2 Gyr) and the last major 
merger (with a mass ratio near 1:1) encountered at $z$=2.2 (age=3 Gyr).  The 
interaction between the merging galaxies lasts for $\sim$1~Gyr and 
significantly disrupts the primary disc, although the disc 
re-establishes itself rapidly due to its gas-rich nature. There is a 
delay between the last major merger and the time at which the polar disc 
becomes noticeable (at redshift $z$=0.8, age=6.8 Gyr). The polar disc then 
establishes itself, evolving passively and stably for the last 
$\sim$4~Gyrs of the simulation. In the \citet{Brook2008} galaxy 
studied here, the polar disc component of the galaxy initially forms as 
a ring but over the following Gyrs the centre is filled by cold gas to 
resemble a disc. For a video of the polar disc formation, first presented in \citet{Brook2008}, see 
http://www.youtube.com/watch?v=c-H3WzaewdY. Only in the final $\sim$0.2~Gyr does 
the polar disc become unstable due to coupling between the two discs, 
and the polar disc structure vanishes when the two align. There is some 
evidence of a bar growing in the polar disc right up until the disc is 
disrupted, and a bar is evident in the old stellar disc as well. The 
surprising longevity of simulated polar disc structures was highlighted 
by both \citet{Maccio2006} and \citet{Brook2008}. The former note 
that only gas which falls into the polar disc at $\sim$90$^\circ$ to the 
massive stellar disc can form a stable secondary disc; at smaller angles 
of incidence, the infalling gas was found to fall rapidly to the already 
existing disc.

\subsection{Star Formation History}

\begin{figure}
\centering
\includegraphics[scale=0.4,clip]{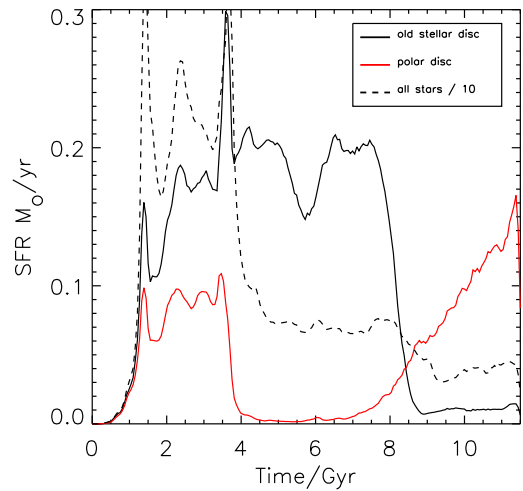}
\caption{Time evolution of the star formation rate of the simulated 
system. The solid black line shows the star formation rate inferred from 
stars presently situated within a slit aligned with the old stellar 
disc and excluding the first kpc; the solid red line corresponds to the star formation rate inferred 
from a slit aligned with the polar disc; the dashed line 
corresponds to the global star formation rate of the system scaled down 
by a factor of ten. This is an expanded version of \citet{Brook2008} figure 2.}
\label{Fig:SFR} 
\end{figure}

The polar disc system experiences a fairly constant, intense, phase of 
star formation between 1$-$4~Gyrs after the simulation commenced 
(Fig~\ref{Fig:SFR}). Subsequent to this, the global star formation 
decreases rapidly for the final $\sim$8~Gyrs of the simulation (dashed 
black curve). We can infer the star formation histories along each disk 
by placing `slits' along each (with the knowledge that by doing so means 
some `duplication' of star formation histories in the region of overlap 
between the slits). Although we use the two discs to define where we 
place the `slits' we include all stars along the `slit' in our determination 
of the SFR. The result of this suggests that the primary stellar disc 
forms its stars preferentially at early times, prior to the last major 
merger, although residual gas does remain to form stars at later times 
(solid black curve).  Over the final $\sim$5~Gyrs of the simulation, 
stars start to form in the polar disc (solid red curve - see also Fig~2 
of \citet{Brook2008}), increasing steadily, while the star formation in 
the primary stellar disk has essentially halted over the final 
$\sim$4~Gyrs. During this era, the star formation in the polar disc is 
very smooth and continuous, with no starburst features indicative of 
merging events. Star formation is ongoing and increasing, up to z=0.17 but 
there is no sign of bursting, with a considerable fraction of stars being 
produced recently, so we expect young stars in the polar disc, as in NCG 4650A 
\citep{Gallagher2002}, although no spiral can be seen such as in NGC 4650A.
Star formation is uniform throughout the disc, so there are no young star cluster
regions, merely global star formation without clumping. This does not, however, mean that
the SFR is similar to UGC 7576 or UGC9796 which \citet{Spavone2011} fit with an exponentially decaying
SFR. Indeed, the SFR in the polar disc is increasing with time.  
The residual star formation not accounted for between 
the dashed curve and the sum of the two solid curves is that associated 
with the spheroid components of the system. In Fig~\ref{Fig:SFR}, the 
dotted line (representing 10\% of the global star formation rate) shows 
that the star formation within the discs (the `in situ' component) is a 
small fraction of the global star formation at early times.

As noted in \S~3, the system was decomposed using a stellar age cut : 
stars formed after $t$$>$9~Gyrs are considered as part of the polar 
disc, and those formed before that are assigned to the stellar disc.


\section{The Halo}

One of the original reasons for postulating the existence of dark matter 
was the circular velocity profiles of galaxy discs. With polar discs, we 
have circular velocities of two orthogonal components of the dark matter 
density profile and so these objects provide an ideal method for 
studying the shape of the dark matter halo, as highlighted by \citet{CombesA1996}
\citet{Casertano1991}, \citet{Schweizer1983}, \citet{Sackett1994} and 
\citet{Iodice2006}; indeed, making use of the Tully-Fisher relation, as 
applied to polar discs, \citep{Iodice2003} suggested that dark matter 
haloes are oblate and flattened along the polar disc because the 
\textsc{HI} gas polar disc was found to have a faster circular velocity 
than the stellar disc.

In calculating the shape of the dark matter halo directly we use the 
method of \citet{Knebe2010}. The shape of the halo is calculated using 
spherical shells; at each radius, $R_i$ we calculate the inertial tensor 
using all particles within that radius. We use the standard inertial 
tensor, as opposed to the reduced tensor, because this technique 
strongly biases the profile in favour of the central region.

The halo shape is calculated, according to the inertial tensor, as

\begin{equation}
I_{i,j} = \sum_n = m_nx_{i,n}x_{i,n},
\end{equation}

\noindent
where $I_{i,j}$ is the inertial tensor, $n$ is the particle id, $i$ and 
$j$ are the x,y,z components of the particle position vector. The axial 
ratios are $(a,b,c) = 
(\sqrt{\lambda_a},\sqrt{\lambda_b},\sqrt{\lambda_c})$ where $a>b>c$ and 
$\lambda_{a,b,c}$ are the eigenvalues of the tensor. The directions of 
the axes are the corresponding eigenvectors, as outlined by 
\citet{BailinSt2005}.

This method easily identifies the direction of the principle axes of the 
best fit ellipse, but introduces a bias in the halo shape. To correct 
this to match halo sphericity based on potential energy contours we must 
adjust the sphericity parameter, $S=c/a$ by,

\begin{equation}
S_{true} = S_{sphere}^{\sqrt{3}},
\end{equation} 

\noindent
where the sphericity is the ratio between the major and minor axes of 
the halo. This relation is based upon the study by \citet{BailinSt2005}

We will compare our inferred halo shape with that of \citet{Maccio2006},
 in addition to the work of \citet{Bailin2005}, who studied the 
alignment of traditional galaxy discs with their dark matter haloes and 
found that in the inner region there was a tight alignment which becomes 
increasingly divergent with distance. Thus, we have the amount of mass 
within a radius R along two orthogonal directions, $M_1$ and $M_2$. 

In order to calculate the axial ratios we imagine two spheres of radii `a' 
and `b', containing the same mass. The ratio of the volumes of these two 
spheres is the same as the ratio $M_2/M_1$. Therefore the axial ratios of 
the halo are $(M_2/M_1)^{1/3}$.

\subsection{Radial Profile}

Using the inertial tensor approach outlined above, in 
Fig~4 we show the sphericity profile of the system's 
dark halo (upper panel) and the direction of the minor axis of the 
inertial tensor (bottom panel) as a function of galactocentric distance. 
We find the system is embedded within a nearly spherical (S=0.9) halo near the 
edge of the stellar disk ($\sim$5~kpc), but which declines in sphericity 
both within the inner kpc and beyond $\sim$10~kpc. This reaches a minimum 
sphericity at 54 kpc or approximately half the virial radius and increases again 
to a final value of 0.67 at R$_{vir}$. The dark halo is prolate at the edge of the 
stellar disc, distinctly oblate at about 15 kpc and then becomes 
increasingly triaxial.  No obvious `imprint' 
of the polar disc is seen in the sphericity profile. 

This sphericity trend is similar to that of the \citet{Maccio2006} galaxy, which
has an inner region of greater sphericity than the outer region. This will be discussed further, later in this section.

In the bottom 
panel, we can see that the direction of the halo's minor axis in the 
innermost region is orientated 90$^\circ$ to the minor axis in the outer 
region. The inner region is flattened against the direction of rotation 
of the stellar disc and the polar disc is aligned to the axis of 
rotation of the stellar disc. The polar disc has been aligned to the 
$z$-axis of the volume and the stellar disc is approximately aligned to the 
$y$-axis. Thus, it is clear that the minor axis of the halo in the 
innermost region is aligned to the stellar disc and in the outer region 
it is aligned to the polar disc.

\begin{figure}
\centering
\includegraphics[scale=0.4,clip]{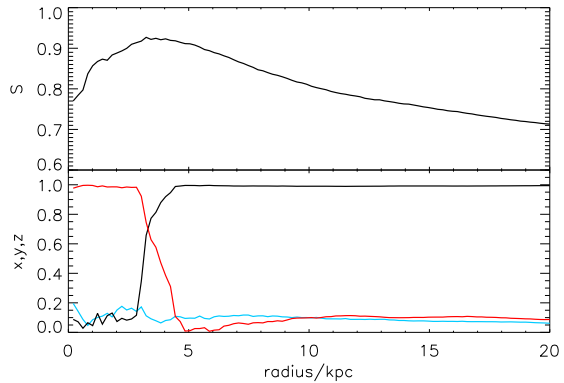}
\label{Fig:fullprofile}
\caption{Upper panel: The sphericity profile of the dark matter halo as 
a function of galactocentric distance. Bottom panel: The direction of 
the minor axis of the inertial tensor (black: $z$-component; red: 
$x$-component; cyan: $y$-component).}
\end{figure}

For non-polar disc galaxies the minor axis of the dark matter halo is 
found to align closely with the angular momentum of the disc (e.g., 
\citet{Bailin2005}. Fig 5 shows an example of 
such a non-polar disc galaxy ({\tt g15784} from \citet{Stinson2010}). 
This galaxy shares some similarities with our polar disc galaxy in that 
it has not experienced a merger or starburst for the past $\sim$4~Gyrs. 
From the bottom panel, one can see that there is no sign of the obvious 
twist in the minor axis profile (cf. Fig 4, for the 
polar disc system here), consistent with the conclusions of Bailin 
et~al, thus demonstrating a significant difference in the properties of 
the two dark haloes.

Another feature of the dark haloes of the \citet{Bailin2005} sample of 
eight galaxies is that although the disc angular momentum and dark halo 
minor axis are tightly aligned in the centre, they diverge with 
increasing radius. The dark halo of the polar disc galaxy seems to 
remain closely aligned at large radii outside the twisted region. Such 
behaviour is seen in only one galaxy of the eight in the Bailin et~al. 
sample. This may be a coincidence, but this strong behaviour in an 
object that is itself uncommon may hold additional clues as to the 
physics by which the polar disc formed. A larger sample size is required 
to exclude the possibility of chance.

\begin{figure}
\centering
\includegraphics[scale=0.4,clip]{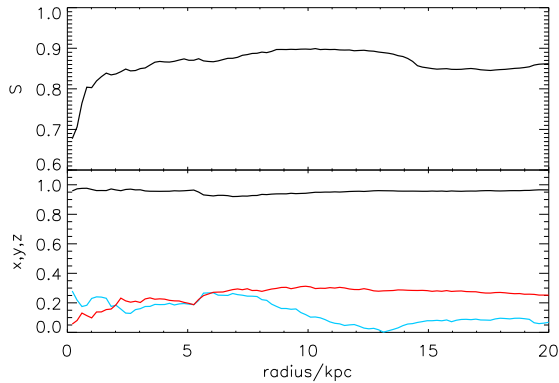}
\label{Fig:darkshapedisc}
\caption{Top panel: The sphericity profile of the dark matter halo as a 
function of galactocentric distance, for a simulated `classical' disc 
galaxy. Bottom panel: The direction of the minor axis of the inertial 
tensor (black: $z$-component; red: $x$-component; cyan: $y$-component).}
\end{figure}

\citet{Iodice2003} note that polar disc galaxies have an offset from 
the Tully-Fisher relation relative to other spiral galaxies, showing 
faster rotation for a given magnitude. This is indicative of flattening 
of the potential towards the plane of the polar disc, as we have 
demonstrated above. Our polar disc galaxy (M$_{\rm I}$=$-$20.3; 
B$-$I=1.7, from \citet{Brook2008} Table 1. This is for the 
whole galaxy as in \citet{Iodice2003}), 
 when compared with the Tully-Fisher relation of S0 galaxies 
\citep{Williams2010} lies above the best-fit line, but not significantly 
so. Considering that \citet{Williams2010} note that the Tully-Fisher 
relation for S0s is lower than that of spirals, the polar disc galaxy 
should actually be indistinguishable from these standard populations, as 
we find when we compare our stellar mass and circular velocity to the results of
\citet{Pizagno2005} and  \citet{McGaugh2005}, among others. This is not unexpected as only 7
of the 16 galaxy sampled by \citet{Iodice2003} are exceptionally displaced from the
general Tully-Fisher profile in their paper. 
A larger sample of simulated polar discs would be required to develop any 
further conclusions in this area, together with a larger sample of K-band observations of PRGs.

\citet{Maccio2006} studied the shape of a simulated polar disc galaxy and calculated
the sphericity of the galaxy halo as we have done here. There are some differences. Their
shape profile peaks at a sphericity of approximately 0.8, at a radius of 9.8 kpc. Our
polar disc peaks at a sphericity of 0.93 at only 3 kpc, corresponding to the edge of
the stellar disc. At half the virial radius our halo is still more spherical than was found by
\citet{Maccio2006} with a sphericity of 0.65 compared to their 0.5. Our values are fairly stable 
with redshift during the lifetime of the polar disc. This galaxy has a sphericity peak 
at approximately twice the inner disc, unlike in our galaxy where 
the peak and the edge of the stellar disc coincide. The sphericity peak in \citet{Maccio2006}
is 0.8 compared to our 0.9 and at half the virial radius the sphericity is 0.5, 
again, less spherical than for our polar disc. Without further comparison in the 
shape finding methods, and a larger statistical sample further conclusions cannot be drawn.

\subsection{Evolution}

\begin{figure}
\centering
\includegraphics[scale=0.45,clip]{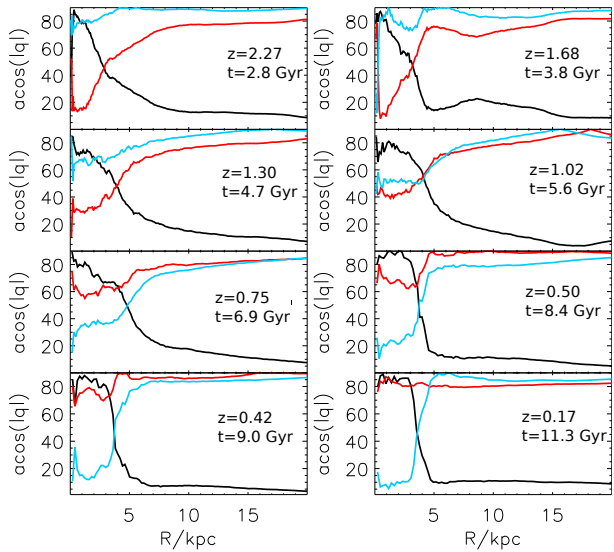}
\label{Fig:shapeevol}
\caption{Time evolution of the alignment profile of the halo's minor 
axis with the simulation aligned to the polar disc. The black line is 
for the $z$-axis (parallel to the polar disc angular momentum), the red 
line for the $x$-axis, and cyan the $y$-axis (approximately parallel to 
the stellar disc angular momentum). The ordinate for each panel is the 
arccosine of the modulus of the projection of the minor axis along each 
of the principle axes. The $y$-axis is defined as the angle of the minor 
axis relative to the $z$-axis of the angular momentum of the cold gas.}
\end{figure}

We analysed both the classical disc galaxy from Fig. \ref{Fig:darkshapedisc} and the polar disc galaxy towards 
the ends of their respective quiescent phases ($\sim$4~Gyrs after the 
last significant merger/burst of star formation, or post-establishment 
of the polar disc, respectively).  The direction of the minor axis of 
the polar disc's dark halo is strongly aligned with the same plane (the 
$z$-axis of the polar disc angular momentum) for much of the simulation 
(Fig.~\ref{Fig:shapeevol}).  This is not the case for the classical 
disc, where the minor axis changes significantly from redshift $z$=0.3 (age=10.3 Gyr)
to $z$=0. Thus, there appears to be a strong, long term alignment in the 
shape and direction of the polar disc halo. This strong alignment may be 
important in giving rise to the polar disc, but we are hindered by our 
sample size of one in coming to a strong conclusion on this point.

The inner region of the polar system shows a long-term `twist' that 
evolves towards orthogonality. The classical disc, after the last 
starburst, sees an evolution towards alignment throughout the halo, such 
that no twist is evident. We suggest that the initial misalignment is 
due to the merger of infalling sub-haloes temporarily re-orienting the 
disc and the alignment thereafter is traced to the later quiescent 
phase. This raises the interesting possibility that alignment of the 
galaxy and dark halo is due to the torques on the galaxy by the dark 
halo. \citet{Bailin2005} discounted this, although they did find that 
the disc is always aligned with the inner halo, which we find to not be 
the case, at least for part of the lives of both discs. For our polar 
disc system, the orientation of the dark matter halo in the outer region 
is set early and does not significantly change over the course of the 
simulation. The classical disc shows considerable variation in the 
direction of the halo minor axis from 0$^\circ$ to 90$^\circ$ over time.

\subsection{Shape from observations}

In order to derive the shape of the polar disc halo from observational 
data alone we can take measurements of the circular velocity of each 
disc. This is best done from the edge-on view because it becomes 
difficult to identify these galaxies simply because of orientation 
effects \citep{Whitmore1990}. By measuring the circular velocity of the 
two discs edge on we can gain insight into the shape of the dark halo.

First, we derive the line-of-sight velocity profiles of the two discs 
by, again, lying down `slits' along both.  These profiles can then be 
used to infer the enclosed mass as a function of galactocentric radius. 
If the inferred mass along one disc is greater than that inferred along 
the orthogonal slit, the former is likely the direction of the major 
axis.  We use the aforementioned redshift $z$=0.17 snapshot (the point 
at which the two discs are near the end of their $\sim$4~Gyr stable 
period and highly orthogonal (86$^\circ$)) to undertake this analysis. 
We arbitrarily assign SPH particles with temperatures T$<$40000K to be 
`cold gas', although our analysis is not sensitive to this choice.  For 
the stellar disc, we concentrate upon stars we know to be rotationally 
supported which were formed between 6$-$6.5~Gyrs after the start of the 
simulation, in order to minimise contamination from the old 
centrally-concentrated disc+bulge stars, the stellar halo, and the polar 
disc itself. Using the circular velocity ($v_{circ}$) and the 
galactocentric radius ($r$), we calculate the mass within that radius 
($M$) as:

\begin{equation}
\label{Eqn:vcirc}
M (<r) = \frac{v_{circ}^2 r}{G},
\end{equation} 
\noindent
where $G$ is the usual gravitational constant.

Using the circular velocity profiles of the two orthogonal discs we can 
probe the dark matter halo using observationally measurable galaxy 
properties. We show the velocity and enclosed mass profiles of the discs 
in Fig~7 (top and bottom panels, respectively). If 
there is more mass within a distance $r$ of the centre of the galaxy 
then the halo is compressed in that direction. From the bottom panel of 
Fig 7, we can see that there is more mass \it 
along \rm the stellar disc than the polar disc, which corresponds to a 
halo compressed orthogonally to the polar disc. At 3~kpc the difference 
in mass is $\sim$70\%, with the mass along the stellar disc being 
greater. Taking the cube root of the mass, to crudely estimate the axial 
ratios, gives 0.9, which matches remarkably well with the direct 
computation of the halo shape, which was found to be 0.92 (Fig. \ref{Fig:fullprofile}),
 confirming the validity of using the two discs 
to probe the shape of the halo.

\begin{figure}
\centering
\includegraphics[scale=0.6,clip]{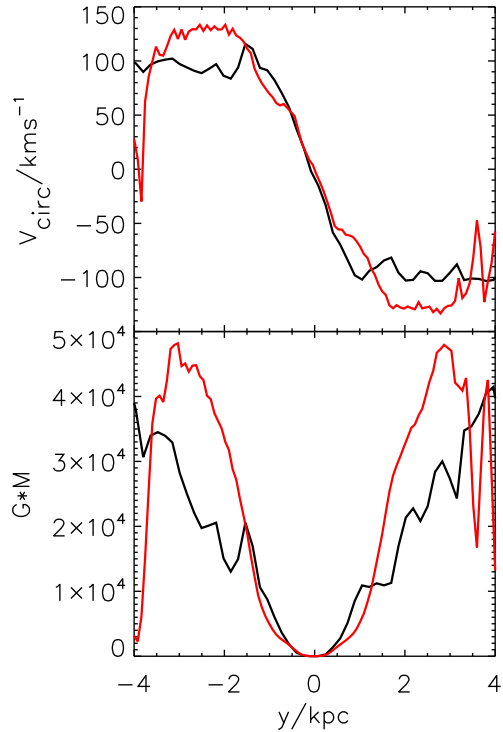}
\label{Fig:circulardisc}
\caption{Top panel: Line-of-sight velocity profile for the intermediate 
age stars (red line) and the polar disc cold gas (black line) at 
redshift $z$=0.17 (age=11.6 Gyr). Bottom panel: Enclosed mass as a function of radius 
based upon eqn~\ref{Eqn:vcirc}.}
\end{figure}

 \citet{Brook2008} note a very high $V/\sigma$ relation for the stellar 
disc of 4.4 and 1.4 for the polar disc. This matches well with the   high $V/\sigma$ 
seen in observations, e.g. \citet{Schweizer1983} who calculate a $V/\sigma$ of 
2.2 for A0136-0801 and \citet{Spavone2011} who give $V/\sigma$ = 1.8, 2.15 and 1.28 
for UGC 7576, UGC 9796 and NGC 4650A respectively.

\section{Formation and Evolution}
\subsection{Angular Momentum}

\begin{figure}
\centering
\includegraphics[scale=0.8,clip]{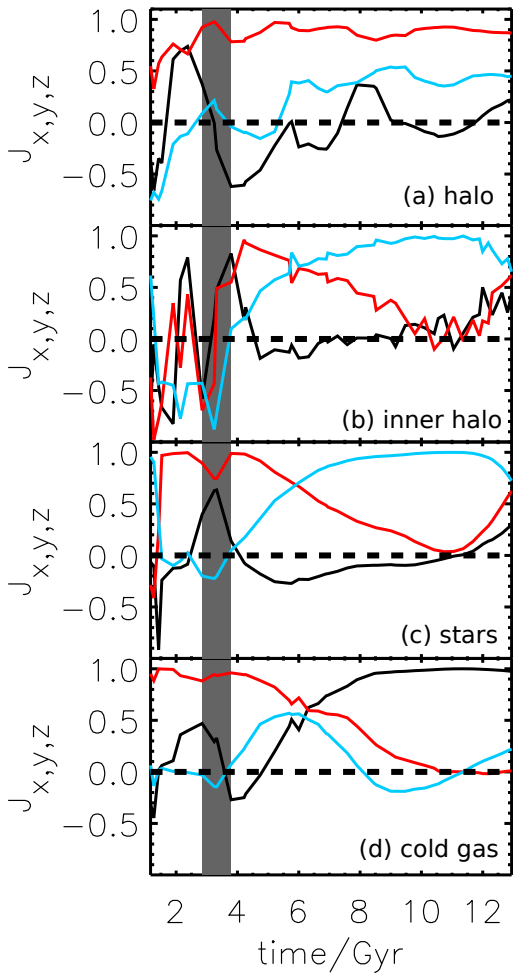}
\label{Fig:angularmomentum}
\caption{Normalised angular momentum projected on the three principle 
axes of the simulation volume : black shows the $z$-axis, red the 
$x$-axis, and blue the $y$-axis. The top panel, labelled `halo', is the 
angular momentum of the entire halo; the panel labelled `inner' is the 
angular momentum of the inner kpc of the halo; the panel labelled 
`stars' is the angular momentum evolution of old ($t$$<$3~Gyrs) stars; 
the panel labelled `cold' shows the trend for the cold gas. The grey 
area is the time of the last major merger.}
\end{figure}

In order to understand the processes that give rise to the polar disc, 
we examine the evolution of the angular momentum of the gas, stars, and 
dark matter. Although not directly related to the disc structures, the 
effect of dark matter is important because of the effect of tidal 
torques.

In order to explore the formation processes of polar disc systems, we 
next traced the origin of the gas that lies within the system and formed 
stars in the discs. \citet{Maccio2006} point out that unless the 
infall of gas is along the polar axis of the potential, it will be 
dragged to align with the stellar disc potential. When gas collapses 
under gravity it retains angular momentum and thus, the gas which 
comprises the polar disc and the stars of the stellar disc at $z$=0.17 
should have a different origin. Two possible origins of the polar disc 
galaxy are suggested: first, the gas that comprises the two discs falls 
along two different trajectories with two different angular momenta, or 
second, the last major merger was inclined to the galaxy and so 
re-oriented the existing gas and stars. New gas would then fall in along 
the old filament, but the early infall baryons would not be re-oriented.

We align the simulation such that the polar disc at redshift $z$=0.17 is 
aligned to the $z$-axis and use that orientation at each earlier 
timestep. The evolution of the angular momentum of the inner and outer 
dark matter halo, the gas, and the stars, is shown in Fig~8.
This plot shows that the alignment of the angular momenta of the old stars, 
intermediate-age stars, gas, and dark matter experience a constant 
evolution with time after 6 Gyr. The polar disc galaxy evolves continuously into its 
$z$=0.17 (11.6 Gyr) position with no sharp transitions.  

The angular momentum of the dark matter halo (panel a) is established 
early and remains essentially constant for the rest of the simulation, 
although some tumbling can be seen in the $z$-axis. The angular 
momentum of the entire halo remains essentially the same throughout 
the history of the simulation, especially the $x$-component. This 
matches with the shape of the dark halo in the outer regions, which 
remains essentially constant.

We see that the angular momentum of the inner region of the dark halo (panel b)
and the stars are locked together because in this region the stars 
dominate the gravitational potential of the halo. 

\begin{figure}
\centering
\includegraphics[scale=0.4,clip]{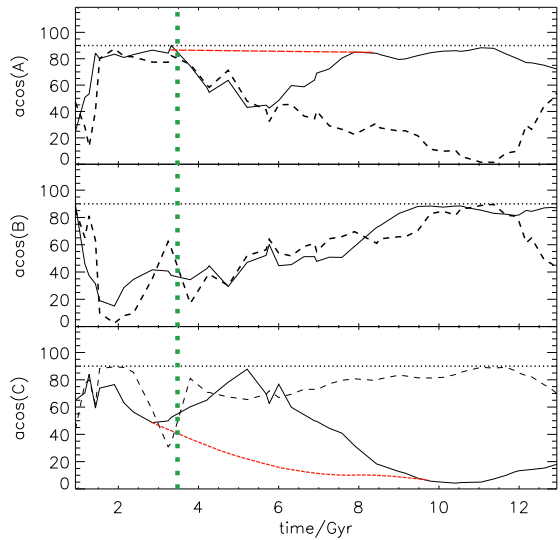}
\label{Fig:crosscorr}
\caption{The angle between the angular momentum of the cold gas (solid 
line) and the old stars (dashed line) and the major (top panel), 
intermediate (middle panel) and minor (bottom panel) axes of the dark 
matter halo. The shape of the halo is measured at 0.25$R_{vir}$. The dotted 
green lines show the time of the last major merger for reference and the 
red lines indicate the angular momentum which the galaxy would presumably have followed
without the last major merger occurring. }
\end{figure}

We have previously discussed that the shape of the polar disc halo, and 
its orientation, remains fairly constant with time. We can use the axes 
of the best-fit ellipse at a radius of 0.25$R_{vir}$ to calculate the 
cross-correlation of the gas and stars with the shape of the halo 
potential, as shown in Fig. \ref{Fig:crosscorr}. 

The angular momentum of the polar disc aligns to the minor 
axis of the dark halo, as previously discussed, and the old stars align 
with the major axis within 2~Gyrs of the last major merger. The cold gas 
takes up to 4~Gyrs to move into close alignment with the minor axis. 
This may be suggestive of interactions between the dark halo potential 
which follows the dark halo shape and the galaxy gas and stars

The old stars (Fig. \ref{Fig:angularmomentum}, panel c) move orthogonally to the cold gas (Fig. \ref{Fig:angularmomentum}, panel d) after the last major merger. 
During the last major merger there is a peak in the $z$-axis angular 
momentum of the stars which appears to have no long-term effect. This can be seen in both Fig. \ref{Fig:angularmomentum}
 and Fig. \ref{Fig:crosscorr} At this point, the 
other two axes begin to twist into their final redshift $z$=0.17 (age=11.6 Gyr)
position.

A striking feature of Fig. \ref{Fig:crosscorr} is that there is no difference 
in the orientation of the gas and stars along the intermediate axis in any given time period (middle panel). 
The cold gas angular momentum becomes increasingly aligned to the halo 
minor axis (bottom panel). The most noticeable feature from this plot is that 
the gas sees a temporary change in direction relative 
to the major axis (top panel) after the major merger, but recovers. The 
red line shows the assumed evolution of the gas angular momentum had the last 
major merger never occurred although this has not been tested. The old stars, 
however, {\emph do not recover} or regain their old angular momentum. 

The
angular momentum of the galaxy at early times is closely
aligned with the intermediate axis of the halo inertial tensor
but becomes increasingly divergent. The gas and stars are
originally closely aligned but diverge after the last major
merger. We see that initially the galaxy angular momentum
is orthogonal to the major and minor axes and aligned to
the intermediate axis. After the last major merger there is a change in the
angle of both components relative to the major axis that the
old stars maintain. The infalling gas returns to orthogonality
within several Gyr. Without the major merger it seems
likely that the gas and stars would remain orthogonal to
the major axis. There is, however, an evolution in the angular
momentum relative to the intermediate axis, such that
the gas and stars angular momentum is initially along this
axis but then increasingly diverges. Had the major merger
not disrupted the system it seems likely that there would be
a continuous evolution of both components until the galaxy
angular momentum aligns to the minor axis in agreement with \citet{Bailin2005} This can be assumed from Fig. \ref{Fig:crosscorr}. The ’recovery
time’ suggested corresponds to a feature of the total angular
momentum of the cold gas. Although not shown in Fig. \ref{Fig:angularmomentum} the last major merger appears to result in an
increase in the total angular momentum of the cold gas and
if followed by a trough which lasts from 4 Gyr to 6 Gyr.
After 6 Gyr the angular momentum of the gas rapidly increases
again. The angular momentum of the old stars does
not change in magnitude during this collision.

\subsection{Formation Process}
The formation process of the galaxy is intimately connected with the evolution
of the angular momentum described in the previous section. This is because the changing
alignment of the galaxy with respect to the large scale environment is what
gives rise to the polar disc properties.

Another manifestation of misaligned gas and stellar discs can be seen in 
the phenomenon of warps. It is perhaps reasonable to assume that warps 
and polar discs might have a common origin. In this section we first briefly examine 
the possibility that a polar disc is an extreme type of warp. We then test the
formation scenario presented in \citet{Brook2008} by tracing the angular 
momentum evolution of the infalling gas, and the direction of the last major merger.

  Warps are 
now thought to be due to the interaction between infalling gas and 
misaligned hot gas (e.g., \citet{Roskar2010}). In our simulated system, 
though, the hot gas is aligned to the polar disc. Warps begin at the 
truncation of the stellar disc and the gas is only misaligned to the 
stellar disc outside this radius \citep{sanchez2009}. The polar disc 
at z=0.17 (11.6 Gyr) is out of alignment at all radii and (essentially)
perfectly orthogonal, suggesting that polar discs and warps are distinct
processes. At first glance, there seems to be no significant similarity
between warps and polar discs. However, there is evolution of the 
direction of the polar disc galaxy, as a whole, after the formation of 
the polar disc (Fig. \ref{Fig:crosscorr}), possibly  does to tidal 
torques. 


We identified gas that comprises the polar disc as all gas that is 
cooler than T=40000K within the inner 10~kpc of the host halo at z=0.17 when the polar disc is most prominent. This 
lower temperature threshold does not change the direction of the angular 
momentum of the gas but results in a thinner, better defined, disc 
structure. We track these particles to their origin, then repeat the 
process tracing the gas that formed stars prior to the last major merger 
at $z$=2. When we plot the positions of these sets of particles there is 
little obvious difference between the two, which one might expect if the 
origin of the gas was different filaments.

If we examine an output from before the last major merger, we see that 
there is a broad range of angular momentum in the infalling gas which 
becomes systematically more aligned to the $z$-axis with time. The 
angular momentum of the infalling gas is surprisingly coherent with 
distance from the galaxy. The total angular momentum of the gas which 
forms the polar disc stars is tightly aligned with the $z$-axis from 
$\sim$4~Gyrs onwards. It is a similar story if we exclude the gas which 
is already in the inner region (within the inner 1~kpc of the galaxy). 
This appears to confirm the \citet{Maccio2006} hypothesis regarding 
the importance of cold flows along filaments for polar disc galaxy 
formation.

Another feature of the polar disc structure is that during intermediate
times the star formation in the old stellar disc is fuelled by newly infalling gas.
The reason for this is that when gas falls into the innermost region of the 
galaxy it is strongly affected by the already formed stars, which changes
the angular momentum of the gas and brings it into alignment with the stellar disc. 
In the inner region the potential is dominated by the stars. An observational confirmation
of this is \citet{CombesA1996} who find that no dark halo is needed to reproduce the
rotation curve of the stellar disc in NGC4650A.

This behaviour no longer occurs after about 9 Gyr. As the entire galaxy is gradually rotating
to bring it into alignment with the galaxy halo shape; it may be that the gas is at a steep
enough angle to the stellar disc so that it is no longer dragged into alignment with the 
stellar disc at later times. This is suggested by \citet{Maccio2006} as to why infalling
gas needs to be approximately orthogonal for a polar disc to be stable.

We also see that a considerable fraction of the gas which forms the 
polar disc is aligned to the z axis when it is still far from the galaxy, but by no means 
all the gas is tightly aligned, and interactions within the galaxy halo are important 
in finally aligning the gas.

The scenario favoured by \citet{Brook2008} suggested that the polar 
disc galaxy is formed if the last major merger has a total angular 
momentum orthogonal to the infalling gas. This interaction re-orientates 
the old gas and stars while the angular momentum of newly infalling gas 
remains unchanged. The coincidence of the last major merger with the 
more unusual behaviour of the angular momentum 
(Fig~8) of the galaxy is suggestive that it is 
essential to forming the galaxy. The mass of stars in the polar disc 
galaxy at $z$=2 is $3.2 \times 10^9$~M$_\odot$ and the merging satellite 
galaxy is $1.5 \times 10^9$~M$_\odot$, making the interaction at least 
1:2.

The angular momentum of the collision is calculated by assuming that 
each galaxy is a point sitting within the reference frame of the 
infalling gas. The relative tangential motion of the collision is 
anti-clockwise. A schematic example of the galaxy-galaxy interaction is 
shown in Fig~10. By examining the internal angular 
momentum of the interacting galaxies we found that most of the angular 
momentum of the interaction is stored in the bulk motion of the galaxies 
rather than in internal motions. The angular momentum of the 
satellite's stars is aligned to the angular momentum of the merger.

The angular momentum of the collision is along the $x$-axis in the chosen reference frame. 
This is orthogonal to the angular momentum of the gas falling into the halo from the filament 
which is aligned to fall in along the  z-axis. As the angular 
momentum of the system must be conserved, we suggest that this provides 
the {\emph initial kick which re-orients the old stars away from the initial 
orientation, see Fig. \ref{Fig:crosscorr}. This figure suggests that the major merger alters the behaviour of the
galaxy away from a matched gas and stellar disc  which reorientates from being aligned to the 
halo intermediate axis to the minor axis.

Much of the gas that forms stars between 4 and 6.5 Gyr has its origin in the 
satellite galaxy involved in the last major merger, and thus contributes 
greatly to the alignment of the stellar disc. Specifically, 40\% of the stars formed
between 4 and 6.5 Gyr were produced from gas which fell in with the last major merger.

The motion of the system 
after the major merger may be due to slow tumbling or torquing by the halo 
the halo as mass continues to be accreted. This is important because the 
stellar disc eventually aligns with the $y$-axis of the system.

\begin{figure}
\centering
\includegraphics[scale=0.85,clip]{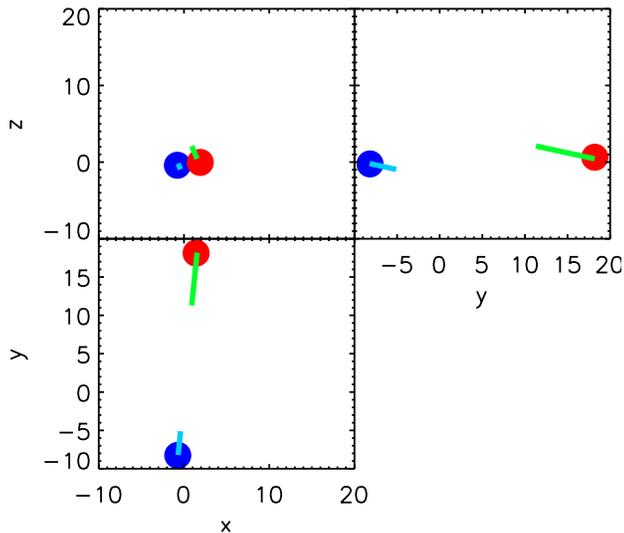}
\label{Fig:collision}
\caption{Projected schematic of the last major merger at redshift $z$=2 (age=3.3Gyr), 
just before the stars merge. The black star is the central galaxy, while 
the red star is the galaxy which merges. The blue and green lines are 
the projected velocity vectors. Distances are in kpcs.}
\end{figure}

\subsection{Metallicity and Age}
\label{sec:metals}

\begin{figure}
\centering
\includegraphics[trim = 0mm 0mm 0mm 0mm,clip,width=80mm]{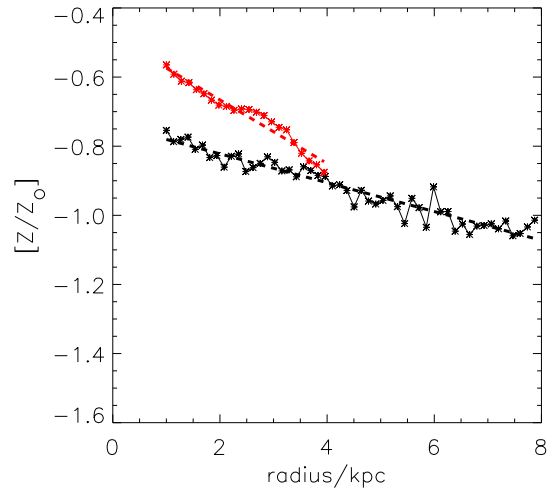} 
\caption{Radial mass-weighted metallicity profile for the stars formed 
in the primary stellar disc (red), and the cold gas of the secondary 
polar disc (black), along 1~kpc thick slits. The stellar disc truncates 
at $\sim$4~kpc; beyond that point, star particles are kinematically 
associated with the stellar halo. The polar disc truncates at 
$\sim$8~kpc. The dotted lines are the best fit lines to the metallicity profiles, 
where $[Z/Z_o]=m*r+c$ and m is the gradient in dex/kpc, and c is the y intercept.} 
\label{Fig:metallicity}
\end{figure}

We derived the radial (mass-weighted) metallicity gradients of the discs 
associated with our simulated polar system by placing narrow slits along 
each component, ignoring the innermost 1~kpc of both, to avoid bulge 
contamination and the intersection point between the two.  The overall 
stellar metallicity of the stellar disk is roughly twice (four times) 
that of the polar stellar (cold gas) disc.

Fig~\ref{Fig:metallicity} shows the radial metallicity profiles for the 
intermediate-age stars of the stellar disk (black) and the cold gas of 
the polar disc (red). The metallicity gradient of the latter is very 
flat ($\sim$$-$0.005$\rightarrow-$-0.01~dex/kpc) consistent with the 
recent observation of flat gradients in NGC~4650A, by 
\citet{Spavone2010}. The metallicity gradients of the stellar and polar 
discs are $-$0.09~dex/kpc and $-$0.04~dex/kpc, respectively. This
matches the results of \citet{Spavone2011}  who find that the 
profile for  UGC 7576 is also extremely shallow.

The negative metallicity gradient seen in Fig~\ref{Fig:metallicity} is 
characteristic of inside-out disc growth \citep{Matteucci1989, 
Matteucci1989a, Boissier1999, Pilkington2011}, consistent with the 
formation scenarios suggested by \citet{Maccio2006} and \citet{Brook2008}
 for their simulated polar disc systems. We did however note that 
the gas in the polar disc initially forms a ring at larger radii and 
then later `fills in' the inner region. This is because while 
the polar disc is forming, the gas in the inner region is inclined to 
the old stellar disc rather than the developing polar disc, depleting the 
inner region of the polar disc structure. We suggest that the abundance 
gradient is indicative of the stellar disc enriching the gas in the 
inner region of the polar disc rather than polar disc stars alone. The 
lower metallicity polar disc is consistent with its younger age sampling 
an earlier part of the age-metallicity relation. It is consistent with a 
polar disc being preferentially `polluted' by the infall of `primordial' 
gas flowing into the system at late times along filaments.  When we 
track the origin of this gas below, we confirm that this is indeed the 
case.

In terms of global metallicity, this polar disc galaxy is similar to
those surveyed by \citet{Spavone2010, Spavone2011}. Specifically, 
Z/Z$_\odot$= 0.2 for our galaxy, matching the metallicity of NGC4650A, 
higher than UGC 9796 (Z/Z$_\odot$=0.1 ) and lower than UGC 7576 (Z/Z$_\odot$=0.4).

The radial age profile of the stars along each disk shows that the stars 
are significantly younger in the polar disc than in the stellar disc 
(Fig~\ref{Fig:stellarage}). The older spheroid component tends to 
dominate at both low and high radii, driving the declines seen at the 
extrema of both curves. If we remove stars that formed in the first 
$\sim$4~Gyrs of the simulation, we find a flat stellar age gradient in 
both the stellar and polar discs. The mean formation time of stars in 
the stellar (polar) disc are 5$-$6~Gyrs ($\sim$10~Gyrs) from the start 
of the simulation.

Looking at Fig. \ref{Fig:SFR} , we see a much higher SFR than is found by
\citet{Spavone2011}. However, their SFR matches that for our 
polar disc at 9~Gyr or about 1~Gyr after it begins to form. This
is a similar age to the estimated age of the polar disc in
NGC 4650A.

\begin{figure}
\centering
\includegraphics[trim = 0mm 0mm 0mm 0mm,clip,width=80mm]{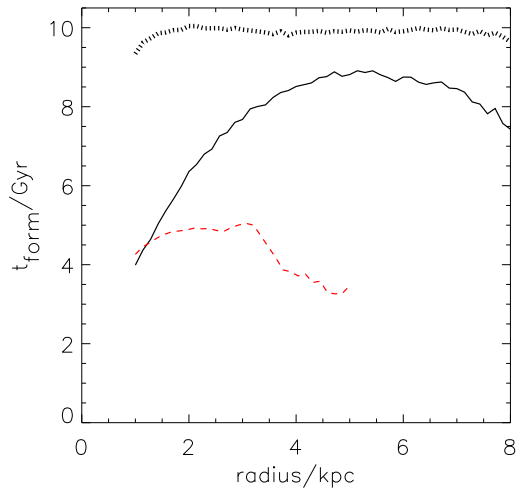} 
\caption{The stellar age profile along the polar (black) and stellar 
(red) discs. $t_{form}$ is the mean formation time of stars at a given 
radius. The stellar (polar) disc truncates at $\sim$4~kpc 
($\sim$8~kpc). The black dotted line shows the star formation time with 
radius for the polar disc, discounting old stars ($t_{form} < 4 Gyr$).}
\label{Fig:stellarage}
\end{figure}

As part of our examination of the temporal evolution of the angular 
momentum, we looked at an alternative approach to decomposing the two 
discs, aligning the simulation volume using the cold gas distribution 
and calculating the angular momentum of each star particle. We can then 
plot the ratio of the $z$- and $y$-components of the angular momentum to 
the total angular momentum, as shown in Fig~\ref{Fig:JzJt}. It is 
noticeable that the angular momentum of stars in the polar disc is 
tightly correlated with the gas, such that all stars with a $J_z/J_t$ 
greater than 0.9 are considered part of the polar disc. For the stellar 
disc, the distribution is broader, suggesting stars with $J_y/J_t$ 
greater than 0.75 are part of the stellar disc.

\begin{figure}
\centering
\includegraphics[trim = 0mm 0mm 0mm 0mm,clip,width=80mm]{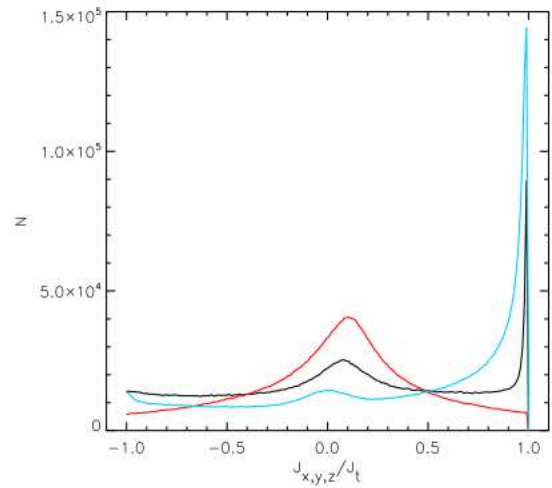} 
\caption{The angular momentum profiles of stars along the $x$ (red), $y$ 
(blue), and $z$ (black) axes, where the cold gas has been aligned with 
the $z$-axis.}
\label{Fig:JzJt}
\end{figure}

The above provides an age-independent method to identify the 
two discs using angular momenta without the age cuts used 
previously. This allows us to see how the discs grow, we bin 
the stars according to their 
distances along their respective discs. If we repeat the calculation of the stellar age profile
using this new method we can see that the slope at radius $<$ 4 kpc and $>$ 6 
kpc is apparently due to contamination from the bulge and stellar halo 
respectively. This shows 
that there is no significant difference in the ages of stars at 
different radii, despite the presence of metallicity gradients.

Fig. \ref{fig:Agebyradius}, panel (a), shows that the star formation
history of the polar disc changes slightly with radius, while panel (b) show the same plot for the older disc. 
In the polar disc, panel (a) shows that there is a population of stars which forms early, contaminating
the sample. Indeed, if we remove those early forming stars ($t_{form} < 4 Gyr$)  from the age profile then the age gradient
is almost completely flat, as if we were using the decomposition method discussed directly above. The black dotted line in Fig 12 shows this trend. In Fig. \ref{Fig:metallicity}, we show the {\it gas} 
metallicity gradient for the polar disc. When we trace the gas back in time, and identify it just before it 
enters the inner region (radius $<$ 12 kpc) we find that the metallicity gradient of the infalling gas is considerably flatter than at z=0.17. 
Additionally, we find no particular gradient in the time at which the gas entered the inner region. This
suggests that the reason for the metallicity gradient at z=0.17, is the enrichment of the gas in the inner region due to 
a higher star formation rate at small radii, which is uniform across time. The metallicity gradient is, apparently, not indicative of an inside-out
formation of the polar disc.

The metallicity distribution of the polar disc stars has the same gradient as the cold gas if we remove the very old stars, but the stars are 0.15 dex less
rich at any given radius. If we include the old stars then we find that the inner region is considerably more metal rich.

We conclude that the \citet{Brook2008} polar disc is 
comparable to the polar disc galaxies observed in nature \citet{Spavone2010, Iodice2002}. The polar disc structure is older than the 
value derived for NGC~4650A by Spavone et~al., but has a comparable mean 
metallicity. The stellar disc is more metal-rich than the younger polar 
disc, and the metallicity gradient for the polar disc is comparably 
shallow to that observed in NGC~4650A. The polar disc does not produce 
stars for $\sim$3$-$4~Gyr after the last major merger, consistent, for 
example, with the picture proposed for UGC~9796 \citep{Cox2006}.

\begin{figure*}
\centering
\subfigure[Polar disc]{
\includegraphics[scale=0.4]{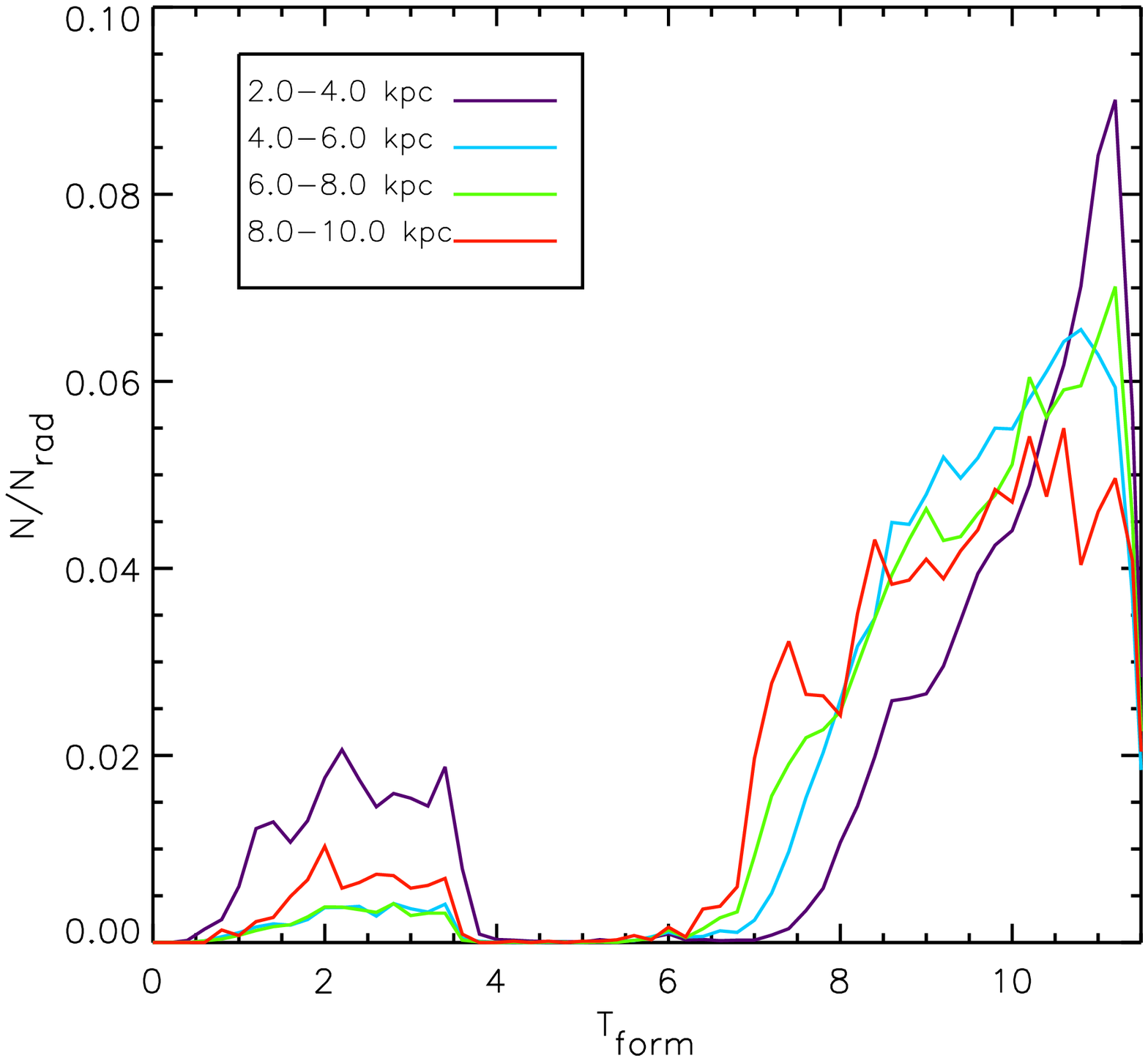}
\label{fig:Agebyradiusa}
}
\subfigure[Stellar disc]{
\includegraphics[scale=0.4]{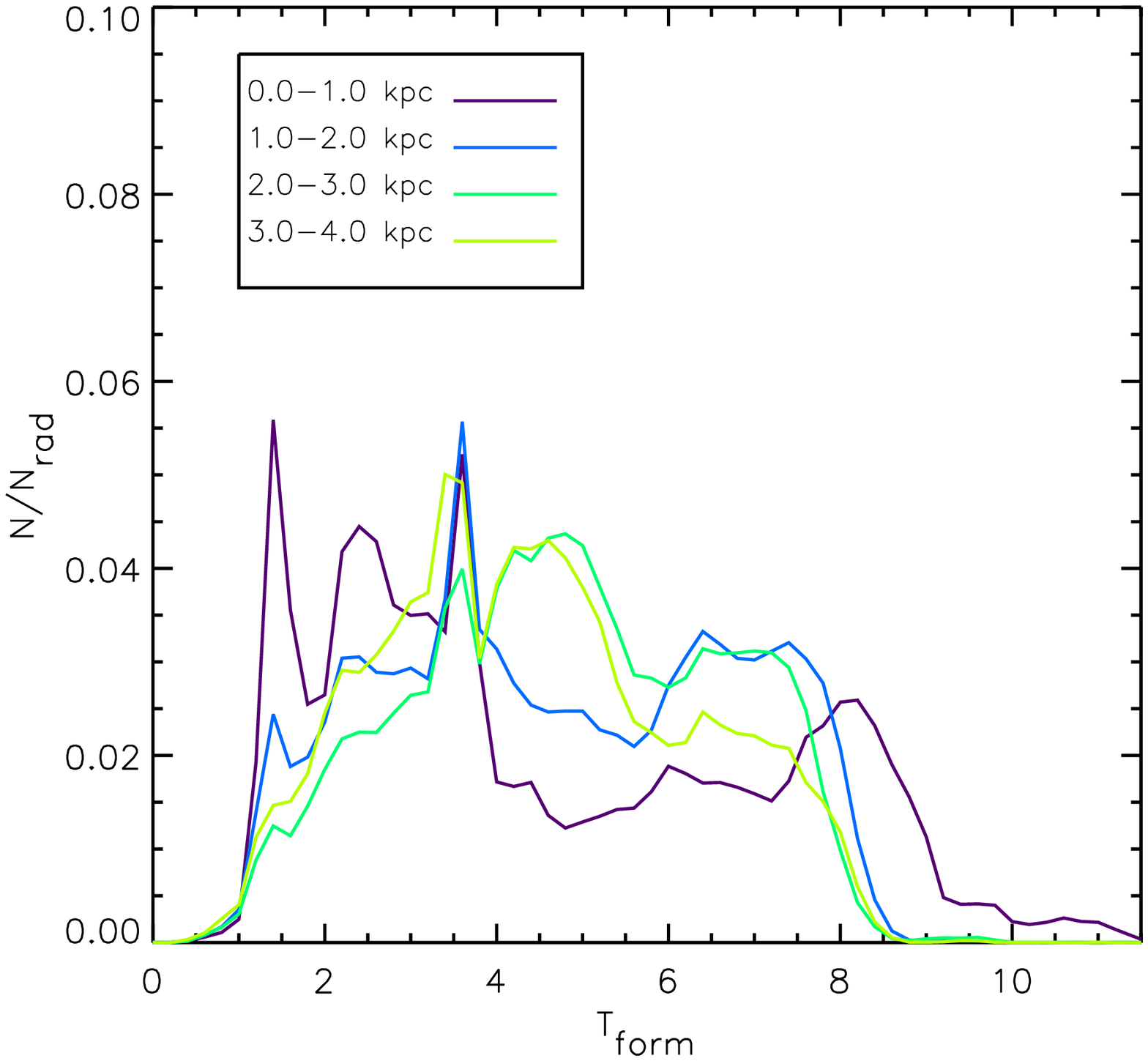}
\label{fig:Agebyradiusb}
}
\label{fig:Agebyradius}
\caption{Age distribution of stars at different radii along 
each disc. The lines are colour-coded as stated in the inset to each
panel.
\subref{fig:Agebyradiusa} shows the polar disc stars, while 
\subref{fig:Agebyradiusb} shows the stellar disc stars }
\end{figure*}
 
\section{Conclusions}

We have extended the preliminary analysis of the polar disc structure 
first introduced by \citet{Brook2008}. We have examined the properties 
of the dark matter halo shape, the formation process of this object and 
the metallicity and stellar age gradients of the discs.

\citet{Iodice2003} point out that polar disc galaxies provide an 
unparalleled chance to study the shape of the inner region dark matter 
haloes using observations in a way normally only possible in simulations. 
The two orthogonal discs have a circular velocity defined by the potential, 
which is dominated by dark matter. By comparing the circular velocity of 
orthogonal discs we estimate the flattening of the dark matter in the 
plane where both discs are seen edge on and find that the halo is
flattened in the direction of the polar disc, as found by Iodice et al. 
Specifically the axial ratio is found to be c/a=0.9. Directly measuring 
the dark matter halo in the simulation we find the same result and compare 
find the axial ratios are extremely similar using both methods, i.e. 
c/a=0.93. The value quoted is for the inner region of the halo, at 
the radius of the polar disc itself. In the outer region of the disc the
sphericity declines to 0.67. We concentrated on the inner region in this work
because that is the region probed by the baryons. 

This value is closely matched 
by simulations of dark matter haloes, such as the galaxy size halos with cooling 
simulated by \citet{Kazantzidis2004} and shown in their figure 2, which shows 
an almost spheroidal inner region, with a sphericity of 0.9 at 10\% if the virial radius,
and an outer region of c/a=0.6, confirming the work of \citet{Springel2004}
and \citet{Allgood2006} amongst others. \citet{Allgood2006}
has categorised halo shapes from cosmological simulations using 
statistical samples. Using their results we expect our halo to have a c/a ratio of 
approximately 0.65 at z=0.17 which is consistent with the polar disc galaxy halo; 
the value of c/a for the `classical' disc shown in Fig. \ref{Fig:darkshapedisc} at the  viral radius 
is 0.7. The shape profile, both in angle and 
sphericity remain stable from outside the inner region to the virial radius. 
This differs from the work of \citet{Bailin2005} and \citet{Allgood2006}
 who find a change of angle and shape with radius. This feature is unusual but 
polar discs are unusual objects and this may be characteristic. \citet{Maccio2006}
did not see the same level of coherence in their polar disc galaxy, however.

Compared to observational estimates, e.g. \citet{Sackett2000}, however, these
halos appear very spherical. Studies of polar ring/disc galaxies (see \citet{Sackett1994},
\citet{CombesA1996}, \citet{Iodice2003}) suggest a highly flattened halo of between 
0.3 and 0.4. \citet{Sackett2000} does, however, provide some suggestions as to why this
difference occurs. Estimates of c/a for halos based on different observations vary wildly 
depending on the chosen metric, such that $c/a = 0.5 \pm 0.2$. In the outer region our
halo falls within this range, 0.65, while in the inner region ($R<15$ kpc) where the 
observational estimates probe we see $c/a$=0.9 to 0.75. The 
simulation outcome is also close to the result of \citet{Hoekstra2004} based 
on weak lensing who estimate a halo sphericity of $0.67^{+0.09}_{-0.07}$, the upper range
of which is close to the sphericity of our halo at a radius of 14 kpc. As for the extremely
elliptical shape found for polar ring/disc galaxies, the methods used may simply not give 
good estimates, indeed, \citet{Sackett2000} notes the difficulty in measuring halo sphericity 
observationally.

We have concluded that the polar disc of this system is not caused directly 
by a merger or stripping. There is no single merger or stripped satellite 
which contains enough gas to form the massive polar disc. By process of 
elimination, the polar disc must form from inflowing cold gas, as in 
\citet{Maccio2006}. We attempted to identify the mechanism by which 
this cold gas flows into the system and gives rise to the polar structure, 
asking, why the filaments should change direction relative to the old stellar 
disc. The structure of the polar disc seems to be a result of the direction 
of gas infall from the filament. We have shown that there is not a single 
moment when the polar disc `forms' but, instead, there is a constant evolution 
of the angular momentum of the infalling gas and stars. 

The angular momentum of the last major merger is almost exactly orthogonal 
to the filament and we feel that this is the most likely origin of the polar 
disc structure. In this scenario, put forwards by \citet{Brook2008}, the 
major merger reorientates the angular momentum of the old stars and the cold 
gas already present in the disc. Gas continues to flow in from the large scale 
structure unaffected by this merger and so builds up the polar disc. 

We see that the major merger affects the stars most strongly without any 
significant influence on the dark halo, and the cold gas recovers its old 
angular momentum within a Gyr. However, this is not the whole story as 
subsequent to the major merger there is further evolution in the orientation 
of the entire galaxy system, potentially due to tumbling or torques from the 
halo. The growing alignment of the polar disc with the intermediate axis of 
the dark halo does not seem affected by the major merger. 

We also provide a basic comparison of the polar disc galaxy to a classical 
disc galaxy to better isolate the origin of the extreme behaviour of the 
polar disc. We also confirm that the polar disc exhibits the 
characteristics of a disc built up from the inside out, both from the 
metallicity gradient and the stellar age profile. Clearly though, any conclusions for their formation, when based 
upon a single simulated system, must be taken as tentative, at best.  We 
need to (and will) repeat this analysis on a statistically significant sample 
of simulated polar disc systems.

\section*{Acknowledgements}
We wish to thank Greg Stinson for supplying the data on the standard, 
none polar, galaxy we used in this paper.
ONS acknowledges the support of STFC through its PhD Studentship 
programme (PPA/S/S/2006/4526) and a fellowship from the European 
Commission’s Framework Programme 7, through the Marie Curie Initial 
Training Network CosmoComp (PITN-GA-2009-238356). BKG, CBB, RJT, and AK 
acknowledge the support of the UK's Science \& Technology Facilities 
Council (ST/F002432/1, ST/G003025/1). BKG acknowledges the generous 
visitor support provided by Saint Mary's University and Monash 
University. AK further acknowledges support by the Spanish Ministerio de 
Ciencia e Innovacion (MICINN) in Spain through the Ramon y Cajal 
programme as well as the grants AYA 2009-13875-C03-02, 
AYA2009-12792-C03-03, CSD2009-00064, and CAM S2009/ESP-1496. 
This work was made possible by the University of Central 
Lancashire's High Performance Computing Facility, the San Diego 
Supercomputing Facility, and the UK's National Cosmology Supercomputer 
(COSMOS). We thank the DEISA consortium, co-funded through EU FP6 
project RI-031513 and the FP7 project RI-222919, for support within the 
DEISA Extreme Computing Initiative. We also thank the referee for their
recommendations which greatly improved the paper. 

\bibliographystyle{mn2e}
\bibliography{PolarDisc}

\begin{thebibliography}{}

\bibitem[\protect\citeauthoryear{{Abadi}, {Navarro}, {Steinmetz} \&
  {Eke}}{{Abadi} et~al.}{2003}]{Abadi2003}
{Abadi} M.~G.,  {Navarro} J.~F.,  {Steinmetz} M.,    {Eke} V.~R.,  2003, \apj,
  591, 499

\bibitem[\protect\citeauthoryear{{Allgood}, {Flores}, {Primack}, {Kravtsov},
  {Wechsler}, {Faltenbacher} \& {Bullock}}{{Allgood}
  et~al.}{2006}]{Allgood2006}
{Allgood} B.,  {Flores} R.~A.,  {Primack} J.~R.,  {Kravtsov} A.~V.,  {Wechsler}
  R.~H.,  {Faltenbacher} A.,    {Bullock} J.~S.,  2006, \mnras, 367, 1781

\bibitem[\protect\citeauthoryear{{Arnaboldi}, {Oosterloo}, {Combes}, {Freeman}
  \& {Koribalski}}{{Arnaboldi} et~al.}{1997}]{Arnaboldi1997}
{Arnaboldi} M.,  {Oosterloo} T.,  {Combes} F.,  {Freeman} K.~C.,
  {Koribalski} B.,  1997, \aj, 113, 585

\bibitem[\protect\citeauthoryear{{Bailin}, {Kawata}, {Gibson}, {Steinmetz},
  {Navarro}, {Brook}, {Gill}, {Ibata}, {Knebe}, {Lewis} \& {Okamoto}}{{Bailin}
  et~al.}{2005}]{Bailin2005}
{Bailin} J.,  {Kawata} D.,  {Gibson} B.~K.,  {Steinmetz} M.,  {Navarro} J.~F.,
  {Brook} C.~B.,  {Gill} S.~P.~D.,  {Ibata} R.~A.,  {Knebe} A.,  {Lewis} G.~F.,
     {Okamoto} T.,  2005, \apjl, 627, L17

\bibitem[\protect\citeauthoryear{{Bailin} \& {Steinmetz}}{{Bailin} \&
  {Steinmetz}}{2005}]{BailinSt2005}
{Bailin} J.,  {Steinmetz} M.,  2005, \apj, 627, 647

\bibitem[\protect\citeauthoryear{{Barnes} \& {Hut}}{{Barnes} \&
  {Hut}}{1986}]{Barnes1986}
{Barnes} J.,  {Hut} P.,  1986, \nat, 324, 446

\bibitem[\protect\citeauthoryear{{Begeman}, {Broeils} \& {Sanders}}{{Begeman}
  et~al.}{1991}]{Begeman1991}
{Begeman} K.~G.,  {Broeils} A.~H.,    {Sanders} R.~H.,  1991, \mnras, 249, 523

\bibitem[\protect\citeauthoryear{{Bekki}}{{Bekki}}{1997}]{Bekki1997}
{Bekki} K.,  1997, \apjl, 490, L37+

\bibitem[\protect\citeauthoryear{{Bekki}}{{Bekki}}{1998}]{Bekki1998}
{Bekki} K.,  1998, \apj, 499, 635

\bibitem[\protect\citeauthoryear{{Boissier} \& {Prantzos}}{{Boissier} \&
  {Prantzos}}{1999}]{Boissier1999}
{Boissier} S.,  {Prantzos} N.,  1999, \mnras, 307, 857

\bibitem[\protect\citeauthoryear{{Bournaud} \& {Combes}}{{Bournaud} \&
  {Combes}}{2003}]{Bournaud2003}
{Bournaud} F.,  {Combes} F.,  2003, \aap, 401, 817

\bibitem[\protect\citeauthoryear{{Brook}, {Governato}, {Quinn}, {Wadsley},
  {Brooks}, {Willman}, {Stilp} \& {Jonsson}}{{Brook} et~al.}{2008}]{Brook2008}
{Brook} C.~B.,  {Governato} F.,  {Quinn} T.,  {Wadsley} J.,  {Brooks} A.~M.,
  {Willman} B.,  {Stilp} A.,    {Jonsson} P.,  2008, \apj, 689, 678

\bibitem[\protect\citeauthoryear{{Casertano} \& {van Gorkom}}{{Casertano} \&
  {van Gorkom}}{1991}]{Casertano1991}
{Casertano} S.,  {van Gorkom} J.~H.,  1991, \aj, 101, 1231

\bibitem[\protect\citeauthoryear{{Combes} \& {Arnaboldi}}{{Combes} \&
  {Arnaboldi}}{1996}]{CombesA1996}
{Combes} F.,  {Arnaboldi} M.,  1996, \aap, 305, 763

\bibitem[\protect\citeauthoryear{{Combes}, {Garc{\'{\i}}a-Burillo}, {Braine},
  {Schinnerer}, {Walter}, {Colina} \& {Gerin}}{{Combes}
  et~al.}{2006}]{Combes2006}
{Combes} F.,  {Garc{\'{\i}}a-Burillo} S.,  {Braine} J.,  {Schinnerer} E.,
  {Walter} F.,  {Colina} L.,    {Gerin} M.,  2006, \aap, 460, L49

\bibitem[\protect\citeauthoryear{{Cox}, {Sparke} \& {van Moorsel}}{{Cox}
  et~al.}{2006}]{Cox2006}
{Cox} A.~L.,  {Sparke} L.~S.,    {van Moorsel} G.,  2006, \aj, 131, 828

\bibitem[\protect\citeauthoryear{{Gallagher}, {Sparke}, {Matthews}, {Frattare},
  {English}, {Kinney}, {Iodice} \& {Arnaboldi}}{{Gallagher}
  et~al.}{2002}]{Gallagher2002}
{Gallagher} J.~S.,  {Sparke} L.~S.,  {Matthews} L.~D.,  {Frattare} L.~M.,
  {English} J.,  {Kinney} A.~L.,  {Iodice} E.,    {Arnaboldi} M.,  2002, \apj,
  568, 199

\bibitem[\protect\citeauthoryear{{Gill}, {Knebe} \& {Gibson}}{{Gill}
  et~al.}{2004}]{Gill2004}
{Gill} S.~P.~D.,  {Knebe} A.,    {Gibson} B.~K.,  2004, \mnras, 351, 399

\bibitem[\protect\citeauthoryear{{Haardt} \& {Madau}}{{Haardt} \&
  {Madau}}{1996}]{Haardt1996}
{Haardt} F.,  {Madau} P.,  1996, \apj, 461, 20

\bibitem[\protect\citeauthoryear{{Hoekstra}, {Yee} \& {Gladders}}{{Hoekstra}
  et~al.}{2004}]{Hoekstra2004}
{Hoekstra} H.,  {Yee} H.~K.~C.,    {Gladders} M.~D.,  2004, \apj, 606, 67

\bibitem[\protect\citeauthoryear{{Huchtmeier}}{{Huchtmeier}}{1997}]{Huchtmeier1997}
{Huchtmeier} W.~K.,  1997, \aap, 319, 401

\bibitem[\protect\citeauthoryear{{Iodice}, {Arnaboldi}, {Bournaud}, {Combes},
  {Sparke}, {van Driel} \& {Capaccioli}}{{Iodice} et~al.}{2003}]{Iodice2003}
{Iodice} E.,  {Arnaboldi} M.,  {Bournaud} F.,  {Combes} F.,  {Sparke} L.~S.,
  {van Driel} W.,    {Capaccioli} M.,  2003, \apj, 585, 730

\bibitem[\protect\citeauthoryear{{Iodice}, {Arnaboldi}, {De Lucia}, {Gallagher}
  III, {Sparke} \& {Freeman}}{{Iodice} et~al.}{2002}]{Iodice2002}
{Iodice} E.,  {Arnaboldi} M.,  {De Lucia} G.,  {Gallagher} III J.~S.,  {Sparke}
  L.~S.,    {Freeman} K.~C.,  2002, \aj, 123, 195

\bibitem[\protect\citeauthoryear{{Iodice}, {Arnaboldi}, {Saglia}, {Sparke},
  {Gerhard}, {Gallagher}, {Combes}, {Bournaud}, {Capaccioli} \&
  {Freeman}}{{Iodice} et~al.}{2006}]{Iodice2006}
{Iodice} E.,  {Arnaboldi} M.,  {Saglia} R.~P.,  {Sparke} L.~S.,  {Gerhard} O.,
  {Gallagher} J.~S.,  {Combes} F.,  {Bournaud} F.,  {Capaccioli} M.,
  {Freeman} K.~C.,  2006, \apj, 643, 200

\bibitem[\protect\citeauthoryear{{Kazantzidis}, {Kravtsov}, {Zentner},
  {Allgood}, {Nagai} \& {Moore}}{{Kazantzidis} et~al.}{2004}]{Kazantzidis2004}
{Kazantzidis} S.,  {Kravtsov} A.~V.,  {Zentner} A.~R.,  {Allgood} B.,  {Nagai}
  D.,    {Moore} B.,  2004, \apjl, 611, L73

\bibitem[\protect\citeauthoryear{{Knebe}, {Libeskind}, {Knollmann}, {Yepes},
  {Gottl{\"o}ber} \& {Hoffman}}{{Knebe} et~al.}{2010}]{Knebe2010}
{Knebe} A.,  {Libeskind} N.~I.,  {Knollmann} S.~R.,  {Yepes} G.,
  {Gottl{\"o}ber} S.,    {Hoffman} Y.,  2010, \mnras, 405, 1119

\bibitem[\protect\citeauthoryear{{Knollmann} \& {Knebe}}{{Knollmann} \&
  {Knebe}}{2009}]{Knollmann2009}
{Knollmann} S.~R.,  {Knebe} A.,  2009, \apjs, 182, 608

\bibitem[\protect\citeauthoryear{{Kroupa}}{{Kroupa}}{2001}]{Kroupa2001}
{Kroupa} P.,  2001, \mnras, 322, 231

\bibitem[\protect\citeauthoryear{{Macci{\`o}}, {Moore} \&
  {Stadel}}{{Macci{\`o}} et~al.}{2006}]{Maccio2006}
{Macci{\`o}} A.~V.,  {Moore} B.,    {Stadel} J.,  2006, \apjl, 636, L25

\bibitem[\protect\citeauthoryear{{Matteucci}, {Franco}, {Francois} \&
  {Treyer}}{{Matteucci} et~al.}{1989}]{Matteucci1989a}
{Matteucci} F.,  {Franco} J.,  {Francois} P.,    {Treyer} M.-A.,  1989, \rmxaa,
  18, 145

\bibitem[\protect\citeauthoryear{{Matteucci} \& {Francois}}{{Matteucci} \&
  {Francois}}{1989}]{Matteucci1989}
{Matteucci} F.,  {Francois} P.,  1989, \mnras, 239, 885

\bibitem[\protect\citeauthoryear{{McGaugh}}{{McGaugh}}{2005}]{McGaugh2005}
{McGaugh} S.~S.,  2005, \apj, 632, 859

\bibitem[\protect\citeauthoryear{{Pilkington}, {Few}, {Gibson}, {Calura},
  {Michel-Dansac}, {Thacker}, {Moll{\'a}}, {Matteucci}, {Rahimi}, {Kawata},
  {Kobayashi}, {Brook}, {Stinson}, {Couchman}, {Bailin} \&
  {Wadsley}}{{Pilkington} et~al.}{2012}]{Pilkington2011}
{Pilkington} K.,  {Few} C.~G.,  {Gibson} B.~K.,  {Calura} F.,  {Michel-Dansac}
  L.,  {Thacker} R.~J.,  {Moll{\'a}} M.,  {Matteucci} F.,  {Rahimi} A.,
  {Kawata} D.,  {Kobayashi} C.,  {Brook} C.~B.,  {Stinson} G.~S.,  {Couchman}
  H.~M.~P.,  {Bailin} J.,    {Wadsley} J.,  2012, \aap, 540, A56

\bibitem[\protect\citeauthoryear{{Pizagno}, {Prada}, {Weinberg}, {Rix},
  {Harbeck}, {Grebel}, {Bell}, {Brinkmann}, {Holtzman} \& {West}}{{Pizagno}
  et~al.}{2005}]{Pizagno2005}
{Pizagno} J.,  {Prada} F.,  {Weinberg} D.~H.,  {Rix} H.-W.,  {Harbeck} D.,
  {Grebel} E.~K.,  {Bell} E.~F.,  {Brinkmann} J.,  {Holtzman} J.,    {West} A.,
   2005, \apj, 633, 844

\bibitem[\protect\citeauthoryear{{Reshetnikov} \& {Sotnikova}}{{Reshetnikov} \&
  {Sotnikova}}{1997}]{Reshetnikov1997}
{Reshetnikov} V.,  {Sotnikova} N.,  1997, \aap, 325, 933

\bibitem[\protect\citeauthoryear{{Ro{\v s}kar}, {Debattista}, {Brooks},
  {Quinn}, {Brook}, {Governato}, {Dalcanton} \& {Wadsley}}{{Ro{\v s}kar}
  et~al.}{2010}]{Roskar2010}
{Ro{\v s}kar} R.,  {Debattista} V.~P.,  {Brooks} A.~M.,  {Quinn} T.~R.,
  {Brook} C.~B.,  {Governato} F.,  {Dalcanton} J.~J.,    {Wadsley} J.,  2010,
  \mnras, 408, 783

\bibitem[\protect\citeauthoryear{{Rubin}}{{Rubin}}{1994}]{Rubin1994}
{Rubin} V.~C.,  1994, \aj, 108, 456

\bibitem[\protect\citeauthoryear{{Sackett}}{{Sackett}}{1999}]{Sackett2000}
{Sackett} P.~D.,  1999, in {D.~R.~Merritt, M.~Valluri, \& J.~A.~Sellwood} ed.,
  Galaxy Dynamics - A Rutgers Symposium Vol.~182 of Astronomical Society of the
  Pacific Conference Series, {The Shape of Dark Matter Halos}.
p.~393

\bibitem[\protect\citeauthoryear{{Sackett}, {Rix}, {Jarvis} \&
  {Freeman}}{{Sackett} et~al.}{1994}]{Sackett1994}
{Sackett} P.~D.,  {Rix} H.,  {Jarvis} B.~J.,    {Freeman} K.~C.,  1994, \apj,
  436, 629

\bibitem[\protect\citeauthoryear{{S{\'a}nchez-Bl{\'a}zquez}, {Courty}, {Gibson}
  \& {Brook}}{{S{\'a}nchez-Bl{\'a}zquez} et~al.}{2009}]{sanchez2009}
{S{\'a}nchez-Bl{\'a}zquez} P.,  {Courty} S.,  {Gibson} B.~K.,    {Brook} C.~B.,
   2009, \mnras, 398, 591

\bibitem[\protect\citeauthoryear{{Schweizer}, {Whitmore} \&
  {Rubin}}{{Schweizer} et~al.}{1983}]{Schweizer1983}
{Schweizer} F.,  {Whitmore} B.~C.,    {Rubin} V.~C.,  1983, \aj, 88, 909

\bibitem[\protect\citeauthoryear{{Seljak} \& {Zaldarriaga}}{{Seljak} \&
  {Zaldarriaga}}{1996}]{Seljak1996}
{Seljak} U.,  {Zaldarriaga} M.,  1996, \apj, 469, 437

\bibitem[\protect\citeauthoryear{{Sparke} \& {Cox}}{{Sparke} \&
  {Cox}}{2000}]{Sparke2000}
{Sparke} L.~S.,  {Cox} A.~L.,  2000, in {F.~Combes, G.~A.~Mamon, \&
  V.~Charmandaris} ed., Dynamics of Galaxies: from the Early Universe to the
  Present Vol.~197 of Astronomical Society of the Pacific Conference Series,
  {New Observations of Polar Ring Galaxies}.
pp 119--+

\bibitem[\protect\citeauthoryear{{Spavone}, {Iodice}, {Arnaboldi}, {Gerhard},
  {Saglia} \& {Longo}}{{Spavone} et~al.}{2010}]{Spavone2010}
{Spavone} M.,  {Iodice} E.,  {Arnaboldi} M.,  {Gerhard} O.,  {Saglia} R.,
  {Longo} G.,  2010, \apj, 714, 1081

\bibitem[\protect\citeauthoryear{{Spavone}, {Iodice}, {Arnaboldi}, {Longo} \&
  {Gerhard}}{{Spavone} et~al.}{2011}]{Spavone2011}
{Spavone} M.,  {Iodice} E.,  {Arnaboldi} M.,  {Longo} G.,    {Gerhard} O.,
  2011, \aap, 531, A21

\bibitem[\protect\citeauthoryear{{Springel}, {White} \& {Hernquist}}{{Springel}
  et~al.}{2004}]{Springel2004}
{Springel} V.,  {White} S.~D.~M.,    {Hernquist} L.,  2004, in {S.~Ryder,
  D.~Pisano, M.~Walker, \& K.~Freeman} ed., Dark Matter in Galaxies Vol.~220 of
  IAU Symposium, {The shapes of simulated dark matter halos}.
p.~421

\bibitem[\protect\citeauthoryear{{Stinson}, {Seth}, {Katz}, {Wadsley},
  {Governato} \& {Quinn}}{{Stinson} et~al.}{2006}]{Stinson2006}
{Stinson} G.,  {Seth} A.,  {Katz} N.,  {Wadsley} J.,  {Governato} F.,
  {Quinn} T.,  2006, \mnras, 373, 1074

\bibitem[\protect\citeauthoryear{{Stinson}, {Bailin}, {Couchman}, {Wadsley},
  {Shen}, {Nickerson}, {Brook} \& {Quinn}}{{Stinson}
  et~al.}{2010}]{Stinson2010}
{Stinson} G.~S.,  {Bailin} J.,  {Couchman} H.,  {Wadsley} J.,  {Shen} S.,
  {Nickerson} S.,  {Brook} C.,    {Quinn} T.,  2010, \mnras, 408, 812

\bibitem[\protect\citeauthoryear{{Swaters} \& {Rubin}}{{Swaters} \&
  {Rubin}}{2003}]{Swaters2003}
{Swaters} R.~A.,  {Rubin} V.~C.,  2003, \apjl, 587, L23

\bibitem[\protect\citeauthoryear{{Tremaine} \& {Yu}}{{Tremaine} \&
  {Yu}}{2000}]{Tremaine2000}
{Tremaine} S.,  {Yu} Q.,  2000, \mnras, 319, 1

\bibitem[\protect\citeauthoryear{{Wadsley}, {Stadel} \& {Quinn}}{{Wadsley}
  et~al.}{2004}]{Wadsley2004}
{Wadsley} J.~W.,  {Stadel} J.,    {Quinn} T.,  2004, \na, 9, 137

\bibitem[\protect\citeauthoryear{{Wechsler}, {Bullock}, {Primack}, {Kravtsov}
  \& {Dekel}}{{Wechsler} et~al.}{2002}]{Wechsler2002}
{Wechsler} R.~H.,  {Bullock} J.~S.,  {Primack} J.~R.,  {Kravtsov} A.~V.,
  {Dekel} A.,  2002, \apj, 568, 52

\bibitem[\protect\citeauthoryear{{Whitmore}, {Lucas}, {McElroy},
  {Steiman-Cameron}, {Sackett} \& {Olling}}{{Whitmore}
  et~al.}{1990}]{Whitmore1990}
{Whitmore} B.~C.,  {Lucas} R.~A.,  {McElroy} D.~B.,  {Steiman-Cameron} T.~Y.,
  {Sackett} P.~D.,    {Olling} R.~P.,  1990, \aj, 100, 1489

\bibitem[\protect\citeauthoryear{{Williams}, {Bureau} \&
  {Cappellari}}{{Williams} et~al.}{2010}]{Williams2010}
{Williams} M.~J.,  {Bureau} M.,    {Cappellari} M.,  2010, \mnras, 409, 1330

\bibitem[\protect\citeauthoryear{{Zaldarriaga}, {Seljak} \&
  {Bertschinger}}{{Zaldarriaga} et~al.}{1998}]{Zald1998}
{Zaldarriaga} M.,  {Seljak} U.,    {Bertschinger} E.,  1998, \apj, 494, 491

\end{thebibliography}

\label{lastpage}

\end{document}